\documentclass[12pt, a4paper]{article}
\usepackage{a4wide,amssymb,amsmath,amscd,}

\def\ba{\begin{array}}
\def\ea{\end{array}}
\def\bc{\begin{center}}
\def\ec{\end{center}}
\def\ben{\begin{enumerate}}
\def\een{\end{enumerate}}
\def\beqa{\begin{eqnarray}}
\def\eeqa{\end{eqnarray}}
\def\beqan{\begin{eqnarray*}}
\def\eeqan{\end{eqnarray*}}
\def\btab{\begin{tabular}}
\def\etab{\end{tabular}}
\def\bit{\begin{itemize}}
\def\eit{\end{itemize}}

\newcommand{\cA}{{\cal A}}

\newcommand{\cH}{{\cal H}}

\newcommand{\cU}{{\cal U}}
\newcommand{\cS}{{\cal S}}

\newcommand{\cF}{{\cal F}}

\newcommand{\T}{{\mathbb T}}
\newcommand{\Z}{{\mathbb Z}}
\newcommand{\R}{{\mathbb R}}
\newcommand{\C}{{\mathbb C}}

\newcommand{\be}{\begin{equation}}
\newcommand{\ee}{\end{equation}}
\newcommand{\bea}{\begin{eqnarray}}
\newcommand{\eea}{\end{eqnarray}}

\newcommand{\vs}{\vspace}

\begin{document}

\topmargin -2pt

\headheight 0pt

\topskip 0mm \addtolength{\baselineskip}{0.20\baselineskip}
\begin{flushright}
UTTG-10-00 \\
{\tt hep-th/0005205}
\end{flushright}

\vspace{10mm}

\begin{center}
{\large \bf Matrix Theory Compactification on Noncommutative
${\mathbb T}^4/{\mathbb Z}_2$}\\

\vspace{7mm}

Eunsang Kim~${}^{a,}$\footnote{eskim@wavelet.hanyang.ac.kr},
Hoil Kim~${}^{b,}$\footnote{hikim@gauss.kyungpook.ac.kr},
and
Chang-Yeong Lee~${}^{c,d,}$\footnote{leecy@zippy.ph.utexas.edu}

\vspace{5mm}
${}^a${\it Department of Industrial and Applied Mathematics,

Kyungpook National University,\\ Taegu 702-701, Korea}\\

${}^b${\it Topology and Geometry Research Center, Kyungpook National University,\\
Taegu 702-701, Korea}\\

${}^c${\it Theory Group, Department of Physics, University of Texas,
Austin, TX 78712, USA} \\ ${}^d${\it
Department of Physics, Sejong University, Seoul 143-747, Korea}\\

\vs{8mm}

\end{center}

\begin{center}
{\bf ABSTRACT}
\end{center}
In this paper, we construct gauge bundles on
a  noncommutative toroidal orbifold ${\T}^4_\theta/{\Z}_2$.
First, we explicitly construct a bundle with constant curvature
connections
on a noncommutative
${\T}^4_\theta$ following Rieffel's method.
Then, applying the appropriate quotient
conditions for its ${\Z}_2$ orbifold,
we find a Connes-Douglas-Schwarz type solution of matrix
theory compactified on ${\T}^4_\theta/{\Z}_2$.
When we consider two copies of a bundle on ${\T}^4_\theta$ 
invariant under the ${\Z}_2$ action,
the resulting Higgs branch moduli space of equivariant constant
curvature connections
becomes
an ordinary
toroidal orbifold  ${\T}^4/{\Z}_2$.


\thispagestyle{empty}

\newpage
\section*{I. Introduction}

The pioneering
work of Connes, Douglas, and Schwarz (CDS)~\cite{cds}
revealing the equivalence between noncommutative
Yang-Mills theory living on the noncommutative torus
and toroidally compactified
IKKT(and also BFSS) M(atrix) theory~\cite{ikkt,bfss}
with the constant 3-form background field
has spurred
various works~\cite{sw} on
noncommutative geometry and M/string theory since then.
It has soon been known that
the T-duality of M(atrix) theory  can be understood
in terms of Morita equivalence of the vector
bundles over noncommutative tori~\cite{sch,rs}.

Many of these works have been related to the torus
compactification and not much has been addressed to the
noncommutative orbifold case.
Recently, Konechny and Schwarz~\cite{kosch} worked out
the compactification of  M(atrix) theory on
the ${\Z}_2$ orbifold of the noncommutative
two torus.
However, physically more relevant compactification on
the ${\Z}_2$ orbifold of noncommutative 4-torus,
 a singular $K_3$ surface,
has not been worked out  so far.
In the commutative case,
systems of D0-branes on the commutative orbifold
${\T}^4/{\Z}_2$ were studied in \cite{ra}, \cite{gly} and it is
our main objective to extend
the result of \cite{ra} to the noncommutative case.

We consider the compactification in the context of IKKT M(atrix)
model~\cite{ikkt} on the orbifold ${\T}^4/{\Z}_2$
where ${\Z}_2$ acts as a central symmetry $x\mapsto
-x$. Thus, we need to
find a Hilbert space $\cH$ and unitary representations of ${\Z}^4$
and ${\Z}_2$ on $\cH$ and Hermitian operators $X$ such that
\begin{align}
&U_iX_jU_i^{-1}=X_j+2\pi \delta^j_i R_i \label{1}\\
&U_iX_\nu U_i^{-1}=X_\nu \label{2}\\
&\Omega X_i\Omega=-X_i\label{3}\\
&\Omega X_\nu \Omega=X_\nu, \text{ \ } \nu=0,5,\cdots,9,\label{4}
\end{align}
Following the
description of \cite{howu} and \cite{kosch}
we can find operator relations compatible with the quotient
conditions (\ref{1})-(\ref{4}):
\begin{align}
&U_iU_j=e^{2\pi i\theta_{ij}}U_jU_i,\label{crs1}\\
&\Omega U_i\Omega =U_i^{-1}, \text{
\ }\Omega^2=1.\label{crs2}
\end{align}
When $\theta=0$, the relations (\ref{crs1}), (\ref{crs2}) describe
a ${\Z}_2$ equivariant vector bundle on the ${\Z}_2$ space ${\T}^4$
and $X_i$ specify an equivariant connection on the bundle. Now the
equivariant version of the Serre-Swan theorem indicates that there
is a one-to-one correspondence between ${\Z}_2$ equivariant
vector bundles on the ${\Z}_2$ space ${\T}^4$ and finitely
generated projective modules over the crossed product
$C^*$-algebra $C({\T}^4)\rtimes_\alpha {\Z}_2$. As a
noncommutative analogue we see that the relations (\ref{crs1}),
(\ref{crs2}) imply that the Hilbert space $\cH$ is simply a module
over the crossed product algebra
$C({\T}^4_\theta)\rtimes_\alpha{\Z}_2$ or
${\cA}_\theta\rtimes_\alpha{\Z}_2$, where $\alpha$ denotes the
action of ${\Z}_2$ on ${\cA}_\theta$ by involution. The crossed
product
${\cA}_\theta\rtimes_\alpha{\Z}_2$ is the $C^*$-completion
of the linear space
of ${\cA}_\theta$-valued functions on ${\Z}_2$.
Thus a general element of
${\cA}_\theta\rtimes_\alpha{\Z}_2$ is a formal linear combinations
of elements of the form $\prod_iU_i^{n_i}\Omega^{\epsilon_i}$,
where $\epsilon_i\in\{0,1\}$. As noted in
\cite{kosch},
a ${\cA}_\theta$-module is a finitely generated
projective module if and only if its corresponding module over
${\cA}_\theta\rtimes_\alpha{\Z}_2$ is finitely generated
projective. Thus,  bundles on a NC torus ${\T}^4_\theta$ is
closely related with bundles on the noncommutative torodial
orbifold ${\T}^4_\theta/{\Z}_2$.

In this paper, we find a projective module solution to the
quotient conditions (\ref{1})-(\ref{4}). First we calculate
a CDS type solution of M(atrix) theory
compactified on the noncommutative 4-torus. There,
we also show explicitly that the dual tori are actually related
to each other through
SO(4,4$|{\Z}$) transformations. From this solution we
discuss that
the
moduli space of constant curvature connections can be identified
with ordinary 4-torus.
Based on such explicit CDS type solution on noncommutative ${\T}^4$,
we find its ${\Z}_2$ orbifold solutions extending the result of
 \cite{ra} to the noncommutative torodial
orbifold ${\T}^4_\theta/{\Z}_2$.

In Section II, we review the 
projective modules over noncommutative
torus.
In Section III, we construct a projective
module on noncommutative 4-torus {\it a la} Rieffel~\cite{rie}
explicitly, and find a CDS type solution of M(atrix) theory
compactified on the noncommutative 4-torus.
It is also shown that
 the dual torus is actually related via
SO(4,4$|{\Z}$) transformation.
In Section IV, we find a solution for the noncommutative toroidal
orbifold. From this solution we study the moduli space of
equivariant constant curvature connections.
We conclude in Section V.

\section*{II. Noncommutative vector bundles over noncommutative torus}\label{nct4}

In this section we review 
noncommutative vector bundles over NC $d$-torus ${\T}^d_\theta$,
following the lines of \cite{cr,rie,sch,rs}. Recall that
${\T}^d_\theta$ is the deformed algebra
of the algebra of smooth functions on the torus ${\T}^d$ with the
deformation parameter $\theta$, which is a real $d\times d$
anti-symmetric matrix. This algebra is generated by operators $U_1,\cdots,U_d$
obeying the following relations
\begin{align*}
U_iU_j=e^{2\pi i \theta_{ij}}U_jU_i \text{ \ and \ }
U_i^*U_i=U_iU_i^*=1, \text{ \ \ } i,j=1,\cdots,d.
\end{align*}
The above relations define the presentation of the involutive
algebra
$${\cA}_\theta^d=
\{\sum a_{i_1\cdots i_d}U_1^{i_1}\cdots U_d^{i_d}\mid a=(a_{i_1\cdots i_d})\in
{\cS}({\Z}^d)\}$$
where ${\cS}({\Z}^d)$ is the Schwartz space of
sequences with rapid decay. According to the dictionary in
\cite{cos}, the construction of a noncommutative vector bundle
over ${\T}^d_\theta$
corresponds to the construction of finitely generated
projective modules over ${\cA}^d_\theta$. It was proved in
\cite{rie} that
every projective module over a smooth algebra
${\cA}^{d}_{\theta}$
can be represented by a direct sum of modules of the form
${\cS}({\R}^p\times{\Z}^q\times F)$, the linear space
of Schwartz functions on ${\R}^p\times{\Z}^q\times F$, where $2p+q=d$
and $F$ is a finite abelian group.
 The module action is specified
by operators on ${\cS}({\R}^p\times{\Z}^q\times F)$ and the
commutation relation of these operators should be matched with
that of elements in ${\cA}^{d}_{\theta}$.

On such bundles or modules there are  notions of connections and the
Chern character
 \cite{cds,sch,rs,con}.
 Recall that there is the dual action of
the torus group ${\T}^d$ on ${\cA}_\theta^d$ which gives a Lie
group homomorphism of ${\T}^d$ into the group of automorphisms of
${\cA}_\theta^d$. Its infinitesimal form generates a homomorphism of
Lie algebra $ L$ of ${\T}^d$ into Lie algebra of derivations of
${\cA}_\theta^d$. Note that the Lie algebra $L$ is abelian and is
isomorphic to ${\R}^d$. Let
$\delta:L\rightarrow {\rm {Der \ }}({\cA}_\theta^d)$ be the homomorphism.
For each $X\in L$, $\delta(X):=\delta_X$ is a
derivation i.e., for $u,v\in {\cA}_\theta^d$,
\begin{align}
\delta_X(uv)=\delta_X(u)v+u\delta_X(v).\notag
\end{align}
Derivations corresponding to the generators $\{e_1,\cdots,e_d\}$ of
$L$ will be
denoted by $\delta_1,\cdots,\delta_d$. For the generators $U_i$'s
of ${\T}_\theta^d$, it has the following property
\begin{align}
\delta_i(U_j)=2\pi i\delta_{ij}\cdot U_j.\notag
\end{align}
If $E$ is a projective ${\cA}_\theta^d$-module,
a connection $\nabla$ on $E$ is
a linear map from $E$ to $E\otimes L^*$ such that for all $X\in
L$,
\begin{align}
\nabla_X(\xi u)=(\nabla_X\xi)u+\xi\delta_X(u),{\rm { \ \ \
}}\xi\in {E}, u\in {\cA}_\theta^d.\notag
\end{align}
It is easy to see that
\begin{align}
[\nabla_i,U_j]=2\pi i\delta_{ij}\cdot U_j.\notag
\end{align}
Furthermore, for an ${\cA}_\theta^d$-valued inner product
$\langle\cdot,\cdot\rangle$
on $E$, if $\nabla$ has the property that
\begin{align}
\langle \nabla_X\xi,\eta\rangle+\langle\xi,\nabla_X\eta\rangle
=\delta_X(\langle\xi,\eta\rangle),\notag
\end{align}
then it is called a Hermitian connection.
The curvature ${\cF}_\nabla$ of a connection $\nabla$ is a 2-form
on $L$ with values in the algebra of endomorphisms of $E$. That
is, for $X,Y\in L$,
$${\cF}_\nabla(X,Y):=[\nabla_X,\nabla_Y]-\nabla_{[X,Y]}.$$
Since $L$ is abelian, we simply have
${\cF}_\nabla(X,Y)=[\nabla_X,\nabla_Y]$.
Denote
by ${\cal E}={\rm {End}}_{{\cA}_\theta}(E)$
the algebra of endomorphisms of $E$.
Note
that if $\nabla$ and $\nabla'$ are two Hermitian connections,
then $\nabla_X-\nabla'_X$ belongs to the algebra $\cal E$. Thus
once we have fixed a  connection $\nabla$, then
every other  connections is of the form
$\nabla+A$, here $A$ is a linear map $L$ into $\cal E$. In
other words, the space of Hermitian connections is an
affine space with vector space consisting of the linear maps from $L$
to $\cal E$ and also the algebra is related with a moduli space of
a certain connections.

We now consider the endomorphisms algebra of a module over
${\cA}_\theta^d$.
Let $\Lambda$ be a lattice in $H=M\times \widehat{M}$, where
$M={\R}^p\times{\Z}^q\times F$ and $\widehat{M}$ is its dual.
Let $T$ be the corresponding embedding map in the sense of \cite{rie}.
 Thus $\Lambda$ is
the image of ${\Z}^d$ under the map $T$ and this determines a
projective module which will be denoted by ${E}_\Lambda$.
Consider the lattice
\begin{align}
\Lambda^\perp:=\{(m,\hat s)\in M\times \widehat{M}\mid \theta((m,\hat
s),(n,\hat t))=\hat t(m)-\hat s(n)\in{\Z},\text{ \ for all }(n,\hat
t)\in \Lambda\}.\notag
\end{align}
  From the definition, it is easy to see that every operator of the
form
$${\cal U}_{(m,\hat s)}=(n)=e^{2\pi i \hat s(n)}f(n+m)$$
for $(m,\hat s)\in\Lambda^\perp$, commutes with all
operators ${\cU}_{(n,\hat t)}$, $(n,\hat t)\in\Lambda$. In fact
one can show that the algebra of endomorphisms on ${E}_\Lambda$,
denoted by $\text{End}_{{\cA}_\theta}({E}_\Lambda)$, is a
$C^*$-algebra which is obtained by $C^*$-completion of the space
spanned by operators ${\cal U}_{(m,\hat s)}$, $(m,\hat
s)\in\Lambda^\perp$. As shown in \cite{rie},
the algebra $\text{End}_{{\cA}_\theta}({E}_\Lambda)$
can be identified with a noncommutative torus
${\cA}_{\widehat{\theta}}$, here $\widehat{\theta}$ is a bilinear
form on $\Lambda^\perp$,i.e.,
${\cA}_{\widehat{\theta}}$ is  Morita equivalent to
${\cA}_\theta$.
Recall that a $C^*$-algebra  $A$ is
said to be (strongly) Morita equivalent to $A'$ if
$A'\cong\text{End}_A(E)$ for some finite projective module
$E$.
In general, as was proved in \cite{rs}, a NC torus
${\cA}_{\tilde{\theta}}$ is  Morita equivalent to
${\cA}_\theta$ if $\theta$ and $\tilde{\theta}$ are related by
$\tilde{\theta}=(A\theta+B)(C\theta+D)^{-1}$,
where
$\begin{pmatrix}
A&B\\C&D\end{pmatrix}\in \text{ SO}(d,d|{\Z}).$

We shall now turn to the description of the Chern character. In
general $K_0({\cA}_\theta^d)$ classifies projective modules over
${\cA}_\theta^d$. In fact the positive cone $K_0^+({\cA}_\theta^d)$
corresponds to genuine projective modules and if $\theta$ is not
rational, $K_0^+({\cA}_\theta^d)$ consists exactly of its elements
of strictly positive trace.
The Chern character of a gauge bundle on a noncommutative torus is
an element in the Grassmann algebra $\wedge^\cdot(L^*)$, where $L$
denotes the Lie algebra of ${\T}^d$ and $L^*$ is the dual vector
space of $L$.
Since there is a lattice $D$ in $L$, we see that there are
elements of $\wedge^\cdot D^*$ which
are integral. Now
the Chern character is the map
${\rm { \ Ch  }}:K_0({\cA}_\theta^d)\rightarrow\wedge^{\rm{ev}}(L^*)$
defined by
\begin{align}
{\rm { \ Ch  }}(E):={\widehat \tau}(e^{\frac{\cF}{2\pi i}})=\sum_{k=0}\frac{1}{(2\pi
i)^k}\frac{{\widehat\tau}({\cF}^k)}{k!},\notag
\end{align}
where $E$ is any gauge bundle and $\cF$ is a curvature of an
arbitrary connection on $E$ and
$\widehat\tau$ is a trace on the algebra of endomorphisms.
In general the Chern character
is integral in the commutative case. This is no longer true for
the noncommutative case. However, in the case of noncommutative torus,
there is an integral element related to the Chern character by the
formula
\begin{align}
{\rm{Ch}}(E)=e^{i(\theta)}\mu(E).\label{Ch}
\end{align}
Here $i(\theta)$ denotes the contraction with the deform parameter
$\theta$ regarded as an element of $\wedge^2L$.
The formula (\ref{Ch}) can be realized as a
noncommutative generalization of Mukai vector.
In particular, $\mu(E)=e^{-i(\theta)}{\rm {Ch}}(E)$ is an integral
element of $\wedge^\cdot(L^*)$ which is related with the Chern
character on the classical torus. Also once  we fix the deformation
parameter, then the Chern character ${\rm {Ch}}(E)$ is completely determined by
its integral part $\mu(E)$.
Note that if the 0th component of the
Chern character or the trace is strictly positive, then the gauge
bundle $E$ belongs to the positive cone of $K_0({\cA}_\theta^d)$
and hence it can be written as a direct sum
of the form ${\cS}({\R}^p\times{\Z}^q\times
F)$,  \cite{rie}.

\section*{III. Compactification on noncommutative ${\mathbb
T}^4$.}\label{compact}
In this section we study the compactification solutions on a noncommutative
4-torus ${\T}^4_\theta$
for the case $e^{2\pi i\theta_{ij}}\ne 1$,
following the guide line in \cite{cds}. After we fix
$U_1,U_2,U_3$ and $U_4$, or a projective module, the general
solution has the form of $X_i=\bar X_i+A_i$, where $\bar X_i$ are
particular solutions and $A_i$ are operators commuting with $U_i$.
Here we consider a projective module of the form
${\cS}({\R}^p\times{\Z}^q)\otimes {\cS}(F)$, where $2p+q=4$. Thus
there are three types of modules over ${\cA}_\theta$ according to
$p=0,1,2$. When $p=0$, it is a free module.
The other two types are of the form
${\cS}({\R}\times{\Z}^2)\otimes {\cS}(F)$ and
${\cS}({\R}^2)\otimes {\cS}(F)$. As is discussed in Section II,
a gauge bundle on ${\T}^4_\theta$ correspond to an
element of positive trace which is the 0th component of the Chern
character  and the Chern character is determined by its integral
part $\mu$. Thus it is natural to start with the construction on
${\cS}(F)$ to describe projective modules. Here
we will only consider the case when $p=2$ which is related with
(4220)-systems with a constant curvature considered in \cite{nct4,
gr}.
Let
$F={\Z}_{M_1}\times{\Z}_{M_2}$, where ${\Z}_{M_i}={\Z}/M_i{\Z}$,
($i=1,2$) and
consider the space
${\C}^{M_1}\otimes{\C}^{M_2}$ as the space of functions on
$C({\Z}_{M_1}\times{\Z}_{M_2})$. For all $M_i\in {\Z}$ and
$N_i\in {\Z}/M_i{\Z}$ such that $M_i$ and $N_i$ are relatively
prime, define operators $W_i$ on $C({\Z}_{M_1}\times{\Z}_{M_2})$
by
\begin{align}
(W_1f)(k_1,k_2)&=f(k_1-N_1,k_2)\nonumber\\
(W_2f)(k_1,k_2)&=\exp(-\frac{2\pi ik_1}{M_1})f(k_1,k_2)\nonumber\\
(W_3f)(k_1,k_2)&=f(k_1,k_2-N_2)\notag\\
(W_4f)(k_1,k_2)&=\exp(-\frac{2\pi
ik_2}{M_2})f(k_1,k_2).\nonumber
\end{align}
The operators satisfy the commutation relation
\begin{align}
W_{1}W_{2}&=\exp(2\pi i\frac{N_1}{M_1})W_2W_1\nonumber\\
W_{3}W_{4}&=\exp(2\pi i\frac{N_2}{M_2})W_4W_3,\notag
\end{align}
otherwise commuting. If we write
$W_iW_j=\exp(2\pi i\psi_{ij})W_jW_i$,
then the antisymmetric $4\times 4$ matrix $\psi=(\psi_{ij})$
is of the form
\begin{align}\label{psi}
\psi=\begin{pmatrix} 0&\frac{N_1}{M_1}&0&0\\
                -\frac{N_1}{M_1}&0&0&0\\
                0&0&0&\frac{N_2}{M_2}\\
                0&0&-\frac{N_2}{M_2}&0\end{pmatrix}.
\end{align}
Let
$T:{\mathbb Z}^4\longrightarrow{\mathbb R}^2\times{\mathbb R}^{2*}$
be an embedding map. Thus its matrix representation
$T=\begin{pmatrix}x_{ij}\end{pmatrix}$, $i,j=1,\cdots,4$,
has nonzero determinant and satisfies $(\wedge^2 T^*)(\omega)=-\gamma$
where $\omega =e_3\wedge e_1+e_4\wedge e_2\in \wedge^2({\Z}^4)$ and $e_i$
are standard basis for ${\Z}^4$. Equivalently, if we consider the
Heisenberg representation of ${\Z}^4$ in a Hilbert space, the
desired operators acting on the space of smooth functions on
${\mathbb R}^2$ are defined by the following form:
\begin{align}
(V_if)(s_1,s_2)=(V_{e_i}f)(s_1,s_2):=\exp(2\pi
i(s_1x_{3i}+s_2x_{4i}))f(s_1+x_{1i},s_2+x_{2i}).\notag
\end{align}
These operators obey the commutation relation
\begin{align}
V_iV_j=e^{-2\pi i\gamma_{ij}}V_jV_i,\notag
\end{align}
where
\begin{align}
\gamma_{ij}=\left|\begin{matrix} x_{1i}&x_{1j}\\
                          x_{3i}&x_{3j}\end{matrix}\right| +
             \left|\begin{matrix} x_{2i}&x_{2j}\\
                          x_{4i}&x_{4j}\end{matrix}\right| .\notag
\end{align}

Since $\gamma$ is a real matrix, the operators $V_i$ act on the
Schwartz space ${\mathcal S}({\mathbb R}^2)$. Now we define
operators $U_i=V_i\otimes W_i$ acting on the space
$E_{T}:={\mathcal S}({\mathbb R}^2)\otimes {\mathbb
C}^{M_1}\otimes{\C}^{M_2}$ as follows
\begin{align*}
(U_1f)(s_1,s_2,k_1,k_2)&=e^{2\pi
                 i(s_1x_{31}+s_2x_{41})}
                 f(s_1+x_{11},s_2+x_{21},k_1-N_1,k_2)\\
(U_2f)(s_1,s_2,k_1,k_2)&=e^{2\pi
                 i(s_1x_{32}+s_2x_{42})}\cdot e^{-\frac{2\pi
                 ik_1}{M_1}}
                 f(s_1+x_{12},s_2+x_{22},k_1,k_2)\\
(U_3f)(s_1,s_2,k_1,k_2)&=e^{2\pi
                 i(s_1x_{33}+s_2x_{43})}
                 f(s_1+x_{13},s_2+x_{23},k_1,k_2-N_2)\\
(U_4f)(s_1,s_2,k_1,k_2)&=e^{2\pi
                 i(s_1x_{34}+s_2x_{44})}\cdot e^{-\frac{2\pi
                 ik_2}{M_2}}
                 f(s_1+x_{14},s_2+x_{24},k_1,k_2) .
\end{align*}
Then it is easy to see that they satisfy
\begin{align*}
U_iU_j=\exp(-2\pi i\gamma_{ij}+2\pi i \psi_{ij})U_jU_i.
\end{align*}
Thus we have solution of (\ref{crs1}) if
$\gamma=\psi-\theta$.

Consider operators $\bar X_i$ acting on
$E_{T}={\mathcal S}({\mathbb R}^2)\otimes {\mathbb
C}^{M_1}\otimes{\C}^{M_2}$
given by
\begin{align}
(\bar X_if)(s_1,s_2,k_1,k_2)&=2\pi iA_i^1s_1f(s_1,s_2,k_1,k_2)+2\pi
iA_i^2s_2f(s_1,s_2,k_1,k_2)\nonumber\\
&-A_i^3\frac{\partial f(s_1,s_2,k_1,k_2)}{\partial
s_1}-
A_i^4\frac{\partial f(s_1,s_2,k_1,k_2)}{\partial s_2}\label{xis},
\end{align}
where $A_i^k$ are any real numbers yet to be determined.
 From the definition of $U_i$ and $\bar X_i$, it is easy to see that
the operators $W_i$ are commute with $\bar X_i$. Suppose that
the operators $\bar X_i$ satisfy the equation
(\ref{1}), i.e.,
$$U_i\bar X_jU_i^{-1}=\bar X_j+2\pi \delta^j_i R_i.$$
By a straightforward calculation, the constant matrix $(A_i^j)$ in
(\ref{xis}) can be obtained as in the following form:
$$\begin{pmatrix}R_i A^j_i\end{pmatrix}T=-i\text{ Id}.$$
Since the inverse matrix of $T$ can be written as
$$
T^{-1}=\frac{1}{\det T}
\begin{pmatrix} (-1)^{i+j}B_{ji}\end{pmatrix},\notag
$$
where $B_{ij}$ is the $(ij)$-minor of the matrix
$T$, we see that
\begin{align}\label{relat}
A_i^k=(-1)^{i+k}\cdot\frac{R_i}{i}\cdot\frac{1}{\det T}\cdot
B_{ki},
\end{align}
and this gives  a particular
solution to the equations (\ref{2}) and (\ref{3}).
It is easy to check that the commutator has of the form
\begin{align}
[\bar X_i,\bar X_j]
=2\pi i\left( \left|\begin{matrix}
A_i^1&A_i^3\\A_j^1&A_j^3\end{matrix}\right|
+\left|\begin{matrix} A_i^2&A_i^4\\A_j^2&A_j^4\end{matrix}\right|\right).\notag
\end{align}
By (\ref{relat}), we have
\begin{align*}
[\bar X_i,\bar X_j]&=-2\pi i\cdot\frac{R_iR_j}{(\det
T)^2}\left\{(-1)^{i+1}(-1)^{j+1}\left|\begin{matrix}
B_{1i}&B_{3i}\\B_{1j}&B_{3j}\end{matrix}\right|+(-1)^i(-1)^j
\left|\begin{matrix}
B_{2i}&B_{4i}\\B_{2j}&B_{4j}\end{matrix}\right|\right\}\\
&=2\pi i(-1)^{i+j+1}\cdot \frac{R_iR_j}{(\det
T)^2}\left\{\left|\begin{matrix}
B_{1i}&B_{3i}\\B_{1j}&B_{3j}\end{matrix}\right|+\left|\begin{matrix}
B_{2i}&B_{4i}\\B_{2j}&B_{4j}\end{matrix}\right|\right\}\\
&=2\pi i(-1)^{i+j+1}
\cdot \frac{R_iR_j}{\det
T}\cdot *\gamma_{ij} .
\end{align*}

Now we should find generators of the set of operators which
commute with $U_i$'s. To find such operators we need to describe
an embedding map which corresponds to the dual lattice of the
lattice defined by the embedding map $T$
as
discussed in Section II. For such a map, let
\begin{align}
S=\begin{pmatrix} 0&0&1&0\\
           0&0&0&1\\
           -1&0&0&0\\
           0&-1&0&0\end{pmatrix} \cdot (T^t)^{-1}
           =\frac{1}{\det T}\begin{pmatrix} B_{31}&-B_{32}&B_{33}&-B_{34}\\
                            -B_{41}&B_{42}&-B_{43}&B_{44}\\
                            -B_{11}&B_{12}&-B_{13}&B_{14}\\
B_{21}&-B_{22}&B_{23}&-B_{24}\end{pmatrix}.\label{sma}
\end{align}
Using the matrix (\ref{sma}), we define operators acting on $E_T$ by
\begin{align}
(Z_1f)(s_1,s_2,k_1,k_2)&=e^{\frac{2\pi
i(-s_1B_{11}+s_2B_{21})}{M_1|T|}}\cdot e^{\frac{2\pi
ib_1k_1}{M_1}}f(s_1+\frac{B_{31}}{M_1 |T|},s_2-\frac{B_{41}}{M_1|T|},k_1,k_2)\notag\\
(Z_2f)(s_1,s_2,k_1,k_2)&=e^{\frac{2\pi
i(s_1B_{12}-s_2B_{22})}{M_1|T|}}
f(s_1-\frac{B_{32}}{M_1|T|},s_2+\frac{B_{42}}{M_1|T|},k_1-1,k_2)\notag\\
(Z_3f)(s_1,s_2,k_1,k_2)&=e^{\frac{2\pi
i(-s_1B_{13}+s_2B_{23})}{M_2|T|}}\cdot e^{\frac{2\pi
ib_2k_1}{M_2}}f(s_1+\frac{B_{33}}{M_2|T|},s_2-\frac{B_{43}}{M_2|T|},k_1,k_2)\notag\\
(Z_4f)(s_1,s_2,k_1,k_2)&=e^{\frac{2\pi
i(s_1B_{14}-s_2B_{24})}{M_2|T|}}
f(s_1-\frac{B_{34}}{M_2|T|},s_2+\frac{B_{44}}{M_2|T|},k_1,k_2-1),\notag
\end{align}
where $|T|=\text{ Pf}(\psi-\theta)$ denotes the determinant of $T$
and $b_1$, $b_2$ are integers such that $a_iM_i+b_iN_i=1$, $a_i$
are also integers. To check the operators $Z_i$ commute with all
$U_j$'s, let $Z_iU_j=e^{2\pi i\lambda_{ij}}U_jZ_i$. Then it is easy to see
that
\begin{align}\label{scom}
\lambda_{ij}=\frac{1}{M_k|T|}\left\{\left|\begin{matrix} x_{1i}&x_{3i}\\
(-1)^{3+j}B_{3j}&(-1)^{1+j}B_{1j}\end{matrix}\right|+\left|\begin{matrix}
x_{2i}&x_{4i}\\
(-1)^{4+j}B_{4j}&(-1)^{2+j}B_{2j}\end{matrix}\right|\right\}-
\delta_{ij}\frac{b_kN_k}{M_k},
\end{align}
where $k=1,2$ depending on $ij$. From the relation (\ref{scom}),
\begin{align}
\lambda_{ij}&=0 \text{ \ \ when }i\ne j\notag\\
\lambda_{ii}&=\frac{1}{M_k}-\frac{b_kN_k}{M_k}=\frac{-a_kM_k}{M_k}=-a_k\in {\Z}.\notag
\end{align}
Thus $Z_i$ commute with all $U_j$'s.

Furthermore the operators satisfy
\begin{align}
Z_iZ_j=e^{2\pi i\hat\theta}Z_jZ_i.\label{zs}
\end{align}
Now $\hat\theta$ can be calculated directly and it is given by
\begin{align*}
\hat\theta_{12}&=\frac{a_1N_2+b_1N_2\theta_{12}+a_1M_2\theta_{34}-b_1M_2
\text{Pf}(\theta)}{M_1M_2\text{Pf}(\psi-\theta)}\\
\hat\theta_{13}&=\frac{\theta_{13}}{M_1M_2\text{Pf}(\psi-\theta)}\\
\hat\theta_{14}&=\frac{\theta_{14}}{M_1M_2\text{Pf}(\psi-\theta)}\\
\hat\theta_{23}&=\frac{\theta_{23}}{M_1M_2\text{Pf}(\psi-\theta)}\\
\hat\theta_{24}&=\frac{\theta_{24}}{M_1M_2\text{Pf}(\psi-\theta)}\\
\hat\theta_{34}&=\frac{a_2N_1+b_2N_1
\theta_{34}+a_2M_1\theta_{12}-b_2M_1\text{Pf}(\theta)}
{M_1M_2\text{Pf}(\psi-\theta)}.
\end{align*}
Also we have
\begin{align}
\hat\theta=(A\theta+B)(N-M\theta)^{-1}\label{so44}
\end{align}
where
\begin{align}
A=\begin{pmatrix}
0&-a_1&0&0\\a_1&0&0&0\\0&0&0&-a_2\\0&0&a_2&0\end{pmatrix},\text{ \
\ }B=\begin{pmatrix}
b_1&0&0&0\\0&b_1&0&0\\0&0&b_2&0\\0&0&0&b_2\end{pmatrix}\notag
\end{align}
and
\begin{align}
N=\begin{pmatrix}
N_1&0&0&0\\0&N_1&0&0\\0&0&N_2&0\\0&0&0&N_2\end{pmatrix}\text{ \ \ }
M=\begin{pmatrix}
0&M_1&0&0\\-M_1&0&0&0\\0&0&0&M_2\\0&0&-M_2&0\end{pmatrix}.\notag
\end{align}
 From the equation (\ref{so44}), we see that $-\theta$ and $\hat\theta$
are related by SO$(4,4|{\Z})$ transformation.

Note that U$(n)$ theory on
${\cA}_{-\theta}$ is equivalent to U$(1)$ theory on
${\cA}_{\hat\theta}$. For U$(1)$ theory the generators $Z_i$ can
be identified with functions on the dual torus:
\begin{align}
Z_j\to e^{i\sigma_j}\notag
\end{align}
where $\sigma_j$ are coordinates of the dual torus such that
\begin{align}
[\sigma_i,\sigma_j]=-2\pi i\hat\theta_{ij}.\notag
\end{align}
Now the general solution of the compactification is given by
\begin{align}
X_i=\bar X_i+\sum_{i_1,\cdots,i_4\in{\Z}}
\Psi_{i_1i_2i_3i_4}Z_1^{i_1}Z_2^{i_2}Z_3^{i_3}Z_4^{i_4},\notag
\end{align}
where the coefficients $\Psi_{i_1i_2i_3i_4}$ are $c$-numbers.

Recall that a connection in a module $E_T$ is determined by a set
of operators $\nabla_1,\cdots,\nabla_4$ in $E_T$ such that
\begin{align}
[\nabla_i,U_j]=2\pi i\delta_{ij}U_j.\notag
\end{align}
 From the definition of $\bar X_i$ given in (\ref{xis}) we have
\begin{align}
[\bar X_i,U_j]=-2\pi\delta_{ij}R_jU_j.\notag
\end{align}
Thus  we see that the special solution $\bar X_i$ is related with
connections by
$\bar X_i=\frac{R_i}{i}
\nabla_i$ and for such connection $\nabla$,
the constant curvature ${\cF}=({\cF}_{ij})$ is
given by
\begin{align}\label{curv}
{\cF}=\gamma^{-1}\cdot \text{Id}_N, \text{ \ \ \ where }N=N_1N_2.
\end{align}
Now the general solution should be identified as
\begin{align}\label{sol4}
X_i=\frac{R_i}{i}\nabla_i+A_i(\sigma_1,\sigma_2,\sigma_3,\sigma_4)
\end{align}
where $A_i$ are gauge fields defined on a noncommutative torus.

Note that from the curvature form (\ref{curv}), it
corresponds to the
$U(N)$ gauge theory with vanishing $su(N)$ curvature. This type of
solutions
has
been studied in \cite{bmz} for noncommutative ${\T}^2$ and in \cite{hv}
for higher torus case.  This was generalized to a nonvanishing
 $su(N)$ curvature case
 in \cite{nct4} and it has been noted that the
analysis for noncommutative tori is the same as that of \cite{gr}
for commutative tori. In fact the above solution has been
described by (4220) system with trivial $SU(N)$ gauge fields in \cite{gr}
and its moduli space can be identified with ${\T}^4$. So we may
expect that the moduli space of constant curvature connections in
noncommutative torus
is
of the same form as in the ordinary torus.

The operators
\begin{align}\label{real}
\widetilde\nabla_j=\frac{i}{R_j} \bar X_j+\alpha_j,\text{ \ \ \
}j=1,\cdots,4,
\end{align}
where $\alpha_j$ is any real number, determine a Hermitian
connection with constant curvature in $E_T$. Furthermore
connections of the form (\ref{sol4}) define a representation on
$L^2({\R}^2,{\C}^{M_1}\otimes{\C}^{M_2})$ of
the Heisenberg commutation relations and from this
one can follow
the same steps in  \cite{cr} to show that connections of the form
(\ref{real}) can be found in each gauge orbits
and two such
connections $\frac{i}{R_j} \bar X_j+\alpha_j$ and $\frac{i}{R_j}
\bar X_j+\mu_j$ are gauge equivalent if and only if
$\alpha_j-\mu_j\in{\Z}$. Thus the moduli space of constant
curvature connections
can be identified with $({\R}/{\Z})^4\cong(S^1)^4\cong{\T}^4$.
In general,
if
we consider a projective module consisting of $n$ copies of
such modules, such as $E_{T_1}\oplus\cdots\oplus E_{T_n}$, where $T_i$
is an embedding, then there is a constant curvature connection on
each summand such that the overall
curvature is given by ${\cF}=\oplus{\cF}_k$,
where ${\cF}_k$ is given as in (\ref{curv}) with the same $\gamma$. 
Thus for a
constant curvature connection on $E$
which breaks a projective module $E$ into
$\oplus_kE_{T_i}$,
block diagonal construction gives the moduli space of the form
$({\T}^4)^n/S_n$, where $S_n$ is the symmetric group.

\section*{IV. Compactification on noncommutative toroidal orbifold
${\T}_\theta/{\Z}_2$}

In this section we find  solutions for the quotient conditions
(\ref{1})-(\ref{4}) along with the projective module actions
(\ref{crs1}) and (\ref{crs2}) via
the compactification solutions on a
noncommutative torus ${\T}^4_\theta$ obtained in Section
III. From this we
find the moduli space of equivariant constant curvature
connections on
noncommutative toroidal orbifold ${\T}^4_\theta/{\Z}_2$.

Consider the module
$E_T:={\cS}({\R}^2)\otimes C({\Z}_{M_1})\otimes C({\Z}_{M_2})$
together with $U_i$'s as
operators acting on it. The general solution for the quotient
conditions has been identified as
\begin{align}\label{solu}
X_j=\frac{R_j}{i}\nabla_j+A_j(\sigma_1,\sigma_2,\sigma_3,\sigma_4),
\text{ \ \ \ } 1\le j\le 4.
\end{align}
To find
solutions for the quotient conditions on the compactified part
we need to solve for $\Omega$ which satisfies
$\Omega U_i\Omega=U_i^{-1}$ and $\Omega^2=1$. Consider
an operator
$\Omega_0$ on $E_T$ defined by
\begin{align}
(\Omega_0f)(s_1,s_2,k_1,k_2)=f(-s_1,-s_2,-k_1,-k_2).\notag
\end{align}
It is easy to see that $\Omega_0U_i\Omega_0U_i=e^{2\pi
i(x_{1i}x_{3i}+x_{2i}x_{4i})}$. By redefining
$U_i\mapsto e^{-\pi i(x_{1i}x_{3i}+x_{2i}x_{4i})}U_i$, we get
$\Omega_0 U_i\Omega_0=U_i^{-1}$ and $\Omega_0^2=1$.
Thus we have a solution for (\ref{crs2}) i.e., $\Omega_0$ together with
$U_i$'s
define a projective module over ${\cA}_\theta\rtimes{\Z}_2$.
As  was indicated in \cite{kosch}, there might be other
${\Z}_2$ actions on the module.
To
get other actions on the module, consider the operators
$Z_i$ defined
in Section III. As for the $U_i$'s, rescale $Z_i$ by
$e^{-\pi i(B_{1i}B_{3i}+B_{2i}B_{4i})}Z_i$ and we get the relation
\begin{align}\label{omz}
\Omega_0Z_i\Omega_0=Z_i^{-1}.
\end{align}
Since $Z_i$ commute with all $U_j$'s, the operators
$\Omega_{n_1\cdots n_4}
=e^{i\phi}\Omega_0Z_1^{n_1}Z_2^{n_2}Z_3^{n_3}Z_4^{n_4}$, ($n_i\in
{\Z}$),
satisfy the equation (\ref{crs2}), where $\phi$ is a phase
which is chosen to get the relation $\Omega^2=1$ and it can be calculated
explicitly by using the commutation relations given in (\ref{zs}).
Now consider the general solution (\ref{solu}) satisfying
(\ref{1}) and (\ref{2}). Recall $\nabla_i=\frac{i}{R_i}\bar
X_i$. For $\bar X_i$, which was defined in (\ref{xis}), it is easy
to verify that $\Omega_0\bar X_i\Omega_0=-\bar X_i$.
But since $\bar X_i$ do not commute with $Z_i$'s, we see that $\Omega_0$
is the unique solution for the equation $\Omega\bar X_i\Omega=-\bar X_i$.
By definition of the functions $A_i$ on the dual torus and by the
relation (\ref{omz}), we have
$\Omega_0A_i(\sigma_1,\sigma_2,\sigma_3,\sigma_4)\Omega_0=
A_i(-\sigma_1,-\sigma_2,-\sigma_3,-\sigma_4)$. Applying $\Omega_0$
to the both sides on the equation (\ref{solu}) we see that
\begin{align}\label{ais}
A_i(-\sigma_1,-\sigma_2,-\sigma_3,-\sigma_4)=
-A_i(\sigma_1,\sigma_2,\sigma_3,\sigma_4),
\end{align}
which implies that the functions $A_i$
are odd functions. If we consider a constant curvature connection $\nabla$
on $E_T$, the functions $A_i$ in (\ref{ais}) can be represented by
a
real constant and
hence it vanishes. In other words the moduli space has no Higgs
branch. Note that this type of solutions has been studied in
\cite{ra} for the ordinary torodial orbifold ${\T}^4/{\Z}_2$ under
the name of {\bf Rep. II}.

In the above representation,
the moduli space of constant curvature connections on $E_T$
over ${\T}_\theta^4$ is not preserved by the
${\Z}_2$ action on $E_T$. So it may be more natural to consider
two copies of $E_T$ which respect the ${\Z}_2$ action
and this corresponds to {\bf Rep. I} of \cite{ra}.
Consider the bundle of the form
$E_T^2=E_T\oplus E_T$ and
define operators acting on $E^2_T$
by
\begin{align}
\Omega=\begin{pmatrix} \Omega_0&0\\0&-\Omega_0\end{pmatrix}, \text{ \ \ and
 \ \ }{\bf U}_i=\begin{pmatrix} U_i&0\\0&U_i\end{pmatrix},\notag
\end{align}
where $\Omega_0$ and $U_i$'s are operators on $E_T$ given as above
and in
Section III. Then it is easy to check that
\begin{align}\label{rep1}
{\bf U}_i{\bf U}_j&=e^{2\pi i \theta_{ij}}{\bf U}_j{\bf U}_i,\nonumber\\
\Omega{\bf U}_i\Omega&={\bf
U}_i^{-1}\text{ \ \ \ and \ \ \ }\Omega^2=1.
\end{align}
Thus the relations (\ref{rep1}) defines a projective module over
${\cA}_\theta\rtimes{\Z}_2$. Since $\bar X_i$ defines a particular
solution, we may write the general solution  on the torus as
follows
\begin{align}
X_i=\bar X_i+\begin{pmatrix}
A_i^{11}&A_i^{12}\\A_i^{21}&A_i^{22}\end{pmatrix}.\notag
\end{align}
Since the matrix $\begin{pmatrix}
A_i^{11}&A_i^{12}\\A_i^{21}&A_i^{22}\end{pmatrix}$ should commute
with all the ${\bf U}_i$'s, each entries $A_i^{jk}$ commute with
$U_i$'s. In other words, the operators $A_i^{jk}$ are generated by
$Z_i$'s. Thus they can be identified with functions on the dual
torus. Now the general solutions should be identified as
\begin{align}\label{gsol}
X_i=\frac{R_i}{i}\nabla_i +\begin{pmatrix} A_i^{11}(\sigma_j)
&A_i^{12}(\sigma_j)\\A_i^{21}(\sigma_j)&A_i^{22}(\sigma_j)\end{pmatrix}.
\end{align}
By applying $\Omega$ we find
\begin{align}
\begin{pmatrix} A_i^{11}(-\sigma_j)
&A_i^{12}(-\sigma_j)\\A_i^{21}(-\sigma_j)&A_i^{22}(-\sigma_j)\end{pmatrix}
=\begin{pmatrix} -A_i^{11}(\sigma_j)
&A_i^{12}(\sigma_j)\\A_i^{21}(\sigma_j)&-A_i^{22}(\sigma_j)\end{pmatrix}.\notag
\end{align}
Note that the diagonal entries of the matrix in (\ref{gsol}) are odd
functions on the dual torus, and this fact will be used in finding the
moduli space below. Meanwhile
the off-diagonal entries are even fuctions of $\sigma$.
Here, the gauge transformation should be invariant under $\Omega$
implementing the ${\Z}_2$ quotient condition.
This implies that the gauge parameter in general should be given
by
$\Lambda = \begin{pmatrix}
\lambda_{ev}^{11}&\lambda_{od}^{12}\\
\lambda_{od}^{21}&\lambda_{ev}^{22}\end{pmatrix}$
where the subscript $ev$ or $od$ indicates an even or odd
function of $\sigma$.
This indicates us that not all the $U(2)$ group acts.
We now consider the constant curvature connection $\nabla$ on
$E_T$ considered in Section III. In this case,
as discussed in  {\bf Rep. II} we have constant gauge field in
(\ref{gsol}). Thus
the diagonal entries vanish and the bundle becomes singular
at the fixed points. For the ordinary case
this has been related to the existence of two-brane charge at
the collapsing two-cycle of the blown-up space~\cite{asp95,asp96,doug}.

Now the solutions of the constant curvature
connection in this case
are given by
\begin{align}
X_i&=\bar X_i+
\begin{pmatrix} 0&A_i^{12}(\sigma_j)\\
{A_i^{12}}^\dagger(\sigma_j)&0\end{pmatrix} .\notag
\end{align}
One of the $A_i$ components can be gauged away by constant
gauge transformation of the type $\begin{pmatrix} \lambda &0\\
0& \hat{\lambda} \end{pmatrix}$ which can be decomposed into
two parts, one propotional to the identity and the other proportional to
 $\sigma_3 = \begin{pmatrix} 1 &0\\
0& -1 \end{pmatrix}$.
Since we are only considering constant gauge transformations in dealing
with the moduli space, the
noncommutativity does not affect the result as in Section III.
The remaining component of $A_i$ has translational symmetry of
the commutative 4-torus. This fact together with
a residual guage symmetry $\sigma_3$ now yields a Higgs branch
moduli space of
constant curvature connections to be
an ordinary torodial orbifold ${\T}^4/{\Z}_2$.

For the uncompactified $X_\nu$ sector, the solution is the
same as in the commutative case~\cite{ra}; the moduli
becomes  ${\R}^5 \times {\R}^5 $ when $A_i=0$, and
when $A_i \ne 0$ the transverse moduli becomes ${\R}^5$ for generic
points in ${\T}^4/{\Z}_2$, and
${\R}^5 \times {\R}^5 $ at the fixed points
in ${\T}^4/{\Z}_2$.
Thus this can be viewed as a fibration over
the Higgs branch of ${\T}^4/{\Z}_2$,
with the fiber ${\R}^5$ at
a generic point and with the fiber ${\R}^5 \times {\R}^5 $ at the
orbifold fixed points as suggested in the commutative case \cite{ra}.

For the ordinary ${\T}^4$, the discussion above corresponds to the
construction of the theory of zero branes on ${\T}^4/{\Z}_2$. We
first considered a T-duality on the covering torus ${\T}^4$ to a
dual torus $\hat{\T}^4$ and then project to $\hat{\T}^4/{\Z}_2$.
So, for $N$ identical D0-branes on ${\T}^4/{\Z}_2$ we need
$2N$ zero branes on ${\T}^4$. This is described by $U(2N)$ gauge
theory and the gauge group is broken down to $U(N)\times U(N)$. In
\cite{ra}, it has been shown that the moduli space of the flat
connections is identified with ${\T}^4/{\Z}_2$. In fact our above
analysis on the moduli space of constant curvature connections is
exactly the same as the one in \cite{ra}.


\section*{V. Conclusion and prospect}

In this paper, we construct a bundle on
noncommutative toroidal orbifold ${\T}_\theta^4/{\Z}_2$.
We start with the construction of a bundle
on noncommutative  ${\T}^4$ {\it a la} Rieffel~\cite{rie} and 
find a CDS type solution of M(atrix) theory
compactified on the noncommutative 4-torus. There,  
we also show explicitly that the dual tori are actually related
to each other through
SO(4,4$|{\Z}$) transformations.
Based on  our explicit CDS type solution on noncommutative ${\T}^4$,
we find  its ${\Z}_2$ orbifold solutions, {\bf Rep. I} and
{\bf Rep. II},    
by looking into the  systems of D0-branes on the covering space
 projected onto their invariant parts under the discrete
symmetry group.
 From the solutions obtained, we study the 
moduli space of
equivariant constant curvature connections. 
The
Higgs branch moduli space
has been identified with the ordinary toroidal orbifold 
in the  {\bf Rep. I} case where we
consider two
copies of a bundle over ${\T}_\theta$  which 
are invariant under
the ${\Z}_2$ action on ${\T}_\theta$.
In the {\bf Rep. II} case, the moduli space has no Higgs branch.
In conclusion, in the noncommutative ${\T}^4/{\Z}_2$ case
 the moduli space has the same form as its commutative counterpart.

In \cite{gr}, the moduli space of D0-branes on commutative ${\T}^4$
 with torons of  $U(N)$ Yang-Mills
theory was given as $({\T}^4)^{p_1}/S_{p_1}
\times ({\T}^4)^{p_2}/S_{p_2}$ where $U(N)$ gauge
group broken down into $U(k_1) \times U(k_2)$ satisfying
$k_1+k_2 =N$, and $p_i=gcd(k_i, m_i)$, $i=1,2$ with fluxes
$m_i$ of $U(k_i)$.
Its extension to the noncommutative case has been recently
studied in \cite{nct4} using the 't Hooft's $SU(N)$ solution of
nontrivial twists~\cite{tht}, and the resulting moduli
space of connections turned out to be of the same form,
$({\T}^4)^{p_1}/S_{p_1}
\times ({\T}^4)^{p_2}/S_{p_2}$.
We expect that the same holds for the noncommutative toroidal
 $ {\Z}_2$ orbifold case.
\\

\noindent
{\bf Note added:} After completion of our paper, a related paper
\cite{ks2} has appeared,
which has some overlap with our paper. Their methodology to get
the relevant moduli spaces is to use the theory of representation
of Heisenberg algebra defined by the commutation relations of a
fixed connection.
On the other hand, our approach is the usual one in that we 
construct a module on ${\T}^4_\theta$  with explicit computation,
 and then consider  the ${\Z}_2$ orbifold condition on this module
 finding the moduli space in the specific cases.

\vspace{5mm}
\noindent
{\Large \bf Acknowledgments}

\vspace{5mm}
This work was supported by
Korea Research Foundation,
Interdisciplinary Research
Project 1998-D00001.
We would like to thank KIAS and APCTP
for their kind hospitality where parts of this work
were done. E. K. and H. K. were also supported in part by BK 21.
 C.-Y. L. was also supported in part by NSF PHY-9511632 in Austin.

\newcommand{\J}[4]{{\sl #1}, { #2}, #3, (#4)}
\newcommand{\andJ}[3]{{\bf #1} (#2) #3}
\newcommand{\AP}{Ann.\ Phys.\ (N.Y.)}
\newcommand{\MPL}{Mod.\ Phys.\ Lett.}
\newcommand{\NP}{Nucl.\ Phys.}
\newcommand{\PL}{Phys.\ Lett.}
\newcommand{\PR}{Phys.\ Rev.}

\newcommand{\PRL}{Phys.\ Rev.\ Lett.}
\newcommand{\CMP}{Comm.\ Math.\ Phys.}
\newcommand{\JMP}{J.\ Math.\ Phys.}
\newcommand{\JHEP}{JHEP}
\newcommand{\PTP}{Prog.\ Theor.\ Phys.}
\newcommand{\ib}{{\it ibid.}}
\newcommand{\hep}[1]{{\tt hep-th/{#1}}}

\end{document}